%% file: BoltzmannEnsembleWithPK.tex
\documentclass[a4paper,10pt]{llncs}

\usepackage[centering, height=650pt, width=430pt]{geometry}
\usepackage{amssymb}
\usepackage{amsmath}
\usepackage{graphicx}
\usepackage[latin1]{inputenc}
\usepackage{verbatim}
\usepackage{fourier}
\usepackage{bold-extra}
\usepackage{url}
\usepackage{relsize}
\usepackage{xspace}
\usepackage{tikz}
\usetikzlibrary{arrows}
\usetikzlibrary{shadows}
\usetikzlibrary{trees}
\usetikzlibrary{arrows}
\usetikzlibrary{decorations.pathmorphing}
\usetikzlibrary{backgrounds}
\usetikzlibrary{positioning}
\usetikzlibrary{fit}
\urldef{\mailsa}\path|yann.ponty@lix.polytechnique.fr|
\urldef{\mailsb}\path|saule@lri.fr|
\newcommand{\keywords}[1]{\par\addvspace\baselineskip
\noindent\keywordname\enspace\ignorespaces#1}

\input{macros}

\begin{document}
\mainmatter

\title{A Combinatorial Framework for Designing (Pseudoknotted) RNA Algorithms}
\titlerunning{Hypergraph Ensemble Algorithms for Pseudoknotted RNAs}

%\begin{abstract}
%Unambiguous dynamic programming schemes allow for the generation of suboptimal solutions.
%We explore this property and apply our results to design a new RNA folding algorithm that
%joins two large classes of Pseudoknots. Furthermore, our abstract framework allow us to
%efficiently compute the average number of occurrence of pseudoknots in the Boltzmann ensemble.
%\end{abstract}

\author{Yann Ponty\inst{1}\thanks{To whom correspondence should be addressed} \and Cédric Saule\inst{2,3}}
%

\institute{LIX, École Polytechnique/CNRS/INRIA AMIB, France \\\mailsa
\and LRI, Université Paris-Sud/XI/INRIA AMIB, France\\
\and Institute for Research in Immunology and Cancer, Montreal, Quebec, Canada  \\\mailsb}%


\maketitle

%\toctitle{Lecture Notes in Computer Science}
%\tocauthor{Authors' Instructions}
% ----------------------------------------------------------------
\begin{abstract}
  We extend an hypergraph representation, introduced by Finkelstein and Roytberg,
  to unify dynamic programming algorithms in the context of RNA folding with pseudoknots.
  Classic applications of RNA dynamic programming (Energy minimization, partition function, base-pair probabilities\ldots)
  are reformulated within this framework, giving rise to very simple algorithms.
  This reformulation allows one to conceptually detach the conformation space/energy model -- captured by
  the hypergraph model -- from the specific application, assuming unambiguity of the decomposition.
  To ensure the latter property, we propose a new combinatorial methodology based on generating functions.
  We extend the set of generic applications by proposing an exact algorithm for extracting generalized moments in weighted distribution, generalizing a prior contribution by
  Miklos and al.
  Finally, we illustrate our full-fledged programme on three exemplary conformation spaces (secondary structures, Akutsu's simple type
  pseudoknots and kissing hairpins). This readily gives sets of algorithms that are either novel or have complexity comparable to
  classic implementations for minimization and Boltzmann ensemble applications of dynamic programming.
\end{abstract}

% ----------------------------------------------------------------
\section{Introduction}
\label{sec:intro}
% Introduction Structure

\ParHead{Motivation.}
  Over the past decades biology as a field has become increasingly aware of the importance and diversity of roles played by
  ribonucleic acids (RNA). In addition to playing house-keeping parts, as initially contemplated by the proteo-centric view of cellular processes,
  RNA is now accepted as a major player of gene regulation mechanisms. For instance silencing activity (miRNAs, siRNAs) or
  multi-stable cis-regulatory elements
  (riboswitches) are currently the subject of many research. Furthermore a recent genome-wide experiment has revealed that a large portion
  of the human genome was subject to transcription into RNA.
  While it is unlikely for all these transcripts to be functional as RNAs, novel classes and roles are currently under investigation.
  Most of the functional roles played by RNA require the RNA to adopt a specific structure to make an interaction possible, hide/exhibit
  an active site or allow for a catalytic action (Ribozymes). Being able to understand and simulate how RNA folds is therefore
  a crucial step toward understanding its function.

\ParHead{Ab initio secondary structure prediction.}
  Initial algorithmic methods for the ab-initio prediction of RNA folding considered a coarse-grain conformation space, the secondary
  structure, where each conformation is defined as a non-crossing subset of admissible base-pairs. This led Nussinov and
  Jacobson~\cite{Nussinov1980} to design a $\Theta(n^3)$ dynamic-programming (DP) algorithm for the base-pair maximization problem.
  Building on a nearest neighbor free-energy model proposed by Tinoco \emph{et al}~\cite{Tinoco1973} and extended by the Turner
  group, Zuker and Stiegler~\cite{Zuker1981} created \MFold, a $\Theta(n^3)$ algorithm for minimizing the free-energy (MFE folding), later
  shown to predict correctly $\sim$73\% of base-pairs on a benchmark of RNAs of length $<700$ nucleotides~\cite{Mathews1999}.
  An independent implementation of the
  algorithm is proposed within the popular \Vienna package maintained by  Hofacker~\cite{Hofacker2003a}. Probabilistic alternatives
  (\SFold~\cite{Ding2005}, \ContraFold~\cite{Do2006} and \CentroidFold~\cite{Hamada2009}) have also recently been proposed with substantial improvement,
  relying on a dynamic programming scheme similar to that of \MFold to traverse the conformation space in polynomial time coupled with some
  postprocessing steps.

\ParHead{Ensemble approaches.} Since the seminal work of McCaskill~\cite{McCaskill1990}, the concept of Boltzmann equilibrium  has been used to
embrace the diversity of folding accessible to an RNA sequence. He showed that the partition function of an RNA -- a weighted sum over the
set of all compatible structures -- could be computed through a simple transposition of the DP scheme used for MFE folding. Coupled with a variant of
  the inside/outside algorithm, this led to an exact computation of base-pairs probabilities in the Boltzmann-weighted ensemble.
  This opened the door for more robust predictions, e.g. for RNAs whose MFE folding is an outlier.
  This intuition was later validated by Mathews~\cite{Mathews2004} who showed that the Boltzmann probability correlated well with the actual presence
  of base-pairs in experimentally-determined structures.
  Ding~\emph{et al}~\cite{Ding2005} pushed this paradigm shift a step further by clustering sets of structures sampled within the
  Boltzmann distribution and computing a consensus, improving on the positive-predictive-value (PPV) of existing algorithms.
  This ensemble view naturally spread toward other applications of DP in Bioinformatics
  (sequence alignement~\cite{Mueckstein2002}, simultaneous alignment and folding~\cite{Harmanci2009}, 3D structural alignement~\cite{Ferre2007}),
  and is increasingly becoming a part of the \emph{algorithmic toolbox} of bioinformaticians.

\ParHead{Pseudoknotted conformations.}
  Although substantially successful in their task, secondary structure prediction algorithms were intrinsically limited in
  by their inability to explore conformations featuring crossing base-pairs. Such motifs, called pseudoknots,
  were initially excluded from the conformation space based on the rationale that their participation to the free-energy would remain limited.
  Furthermore, the adjunction of all possible pseudoknots was shown to turn MFE
  folding into an NP-complete problem even in a simple nearest-neighbor model~\cite{Akutsu2000,Lyngsoe2000}.
  However such conformations do naturally occur, and can be essential to functional mechanisms such as -1-frameshift recoding events~\cite{Bekaert2003}
  or the formation of tertiary motifs~\cite{Parisien2008}.
  Therefore many exact DP approaches
  \cite{Rivas1999,Lyngsoe2000,Dirks2003,Reeder2004,Cao2006,Cao2009,Chen2009,Cao2009,Huang2009,Theis2010,Reidys2011}
  have been proposed over the years to extract the MFE structure within  restricted -- polynomially solvable -- classes of pseudoknots.
  However most of these approaches (with the notable exceptions of \cite{Dirks2003,Cao2006,Reidys2011}) were based on ambiguous DP
  schemes, leading them to consider certain structures multiple times. While such an unambiguity would not
  be worrisome in the context of energy minimization, it prevents a direct transposition of these algorithms to ensemble applications
  (partition function, base-pair probabilities) by heavily biasing --  for no biologically valid reason -- derived estimates.

\ParHead{Unambiguous decompositions.}
  This lack of focus on unambiguity in the design of RNA (pseudoknotted) DP algorithms can be explained by two main reasons.
  Firstly certain conformation spaces may not admit unambiguous schemes. Indeed it has been shown by Condon~\emph{et al}~\cite{Condon2004}
  that many PK conformational spaces can be modeled as a formal language, while Flajolet~\cite{Flajolet1987} had shown, using a combinatorial argument, that
  certain simple context-free languages are inherently ambiguous, i.e. not generated by any unambiguous context-free grammar.
  A second explanation is more historical:
  DP algorithms designers were initially focused on optimization problems, and considered the DP equation, not the decomposition of the search space,
  as the central object of their contributions. Indeed in the optimization perspective, it is not mandatory for the conformation space to be
  completely (e.g. sparsification) or unambiguously (e.g. multiply occurring best structure) generated. As decompositions grow more and more complex to
  capture more complex energy models and topological limitations, these two key properties are becoming increasingly hard to ascertain at the level
  of DP equations. Consequently there is a need for more rational framework to facilitate the design of conformational spaces.

  \ParHead{Combinatorial dynamic programming.}
    Over the last century, enumerative combinatorics as a field has been focusing on providing elegant decompositions for all sorts of objects.
    Our proposal is to adopt a similar discipline in the design of DP decompositions, the only task worthy of human attention to our opinion, and will
    eventually lead to an automated procedure for the actual production of codes/algorithms.  To that purpose we chose to
    build on and revisit an hypergraph analogy proposed by Finkelstein~\emph{et al}~\cite{Finkelstein1993} as a unifying framework for RNA folding and other
    applications of DP in Bioinformatics, which we generalize into combinatorial classes amenable to analysis using generating functions.

  \ParHead{Related work.}
    The two main frameworks offering  abstracts view over Dynamic Programming are Lefebvre's multi-tape attributed grammars~\cite{Lefebvre1996}
    and Giegerich's Algebraic Dynamic Programming (ADP)~\cite{Giegerich2000}, respectively building on multitape-attributed grammars and context-free grammars.
    Although very elegant and mature in their implementations, they suffer from limitations in expressivity that are intrinsic to their underlying formalisms. For instance, ADP has to resort to an explicit manipulation of indices in order to achieve competitive complexities
    for canonical pseudoknots~\cite{Reeder2004}, while Lefebvre's multi-tape grammars~\cite{Lefebvre1997} require increased complexity to capture pseudoknots.
    Another formal description of pseudoknotted search spaces is M. Möhl's \emph{split-types}~\cite{Moehl2010}, which
    focuses on how non-contiguous portions are combined, providing a very compact description for pseudoknotted conformation spaces.
    Compared to these abstract representations, the hypergraph formalism achieves a greater expressivity by: i) Implementing an unordered product;
    ii) Allowing explicit manipulation of indices; iii) Allowing additional information to be stored within nodes (Remember that context-free grammars allow for a finite number of non-terminals). For instance,
    polynomial hypergraphs could be proposed for counting homogeneous alignments~\cite{Kucherov2004} whereas these objects cannot be generated by any context-free
    grammar~\cite{Bousquet-Melou2008} and will not be expressed strictly within the alternative frameworks. This improved expressivity comes at a price since the
    manual manipulation of indices is error-prone, as pointed accurately by Giegerich et al, so one may want to think of our proposal as more of a byte code, possibly produced from a higher-level source code (ADP, split-types\ldots).

\ParHead{Outline.}
   In Section~\ref{sec:notations}, we briefly remind some basic definitions related to forward directed hypergraphs. In Section~\ref{sec:generic},
   we remind and propose dynamic programming algorithms for generic problems on F-graphs. Then in Section~\ref{sec:conformations}, we illustrate our programme by
   proposing and proving unambiguous decompositions for three space of conformations: Classic secondary structures in the Turner energy model~\cite{Markham2008a},
   (weighted) base-pair maximisation version of Akutsu's simple-type pseudoknots~\cite{Akutsu2000} and fully-recursive kissing hairpins (Unambiguous restriction
   of Chen~\emph{et al}~\cite{Chen2009}). We also describe a simplified proof strategy based on generating functions to prove the correctness of a given
   decomposition. Section~\ref{sec:moments}
   enriches the scope of applications of our framework by proposing a general algorithm for extracting the moments of additive features (free-energy, base-pairs, helices\ldots) in a weighted distribution (generalizing a previous
   contribution by Miklos~\emph{et al}~\cite{Miklos2005}). Finally Section~\ref{sec:conclu} concludes with some remarks and possible extensions and improvements.

\section{Notations and key notions}\label{sec:notations}
  \begin{figure}[t]
    \begin{center}

% Yann: Balaji, here is my suggestion
      % A Valid F Graph
      \newcommand{\HSpace}{45pt}
      \newcommand{\VSpace}{15pt}
      \newcommand{\VInitTermSpace}{10pt}
      \scalebox{0.8}{
      \begin{tabular}{cc}
      % An Invalid F Graph
      %\tikzstyle{arc}=[->,draw, thick, color=gray]
      \begin{tikzpicture}
       %\pgftext{An Invalid F-Graph}

        % Vertices, placed relatively
        \node[vertex] (n1) {1};
        \node[vertex,right=\HSpace of n1] (n2) {2};
        \node[vertex,right=\HSpace of n2,yshift=-2.2em] (n4) {4};
        \node[vertex,below=\VSpace of n2] (n5) {5};
        \node[vertex,above=\VSpace of n4] (n3) {3};
        \node[vertex,right=\HSpace of n3,yshift=-2.3em] (n6) {6};

        % Init and terminal states
        \node[left=\VInitTermSpace of n1] (i1) {};
        \node[right=\VInitTermSpace of n5] (t5) {};
        \node[right=\VInitTermSpace of n6] (t6) {};

        % Arcs
        \path (n1) edge[arc] (n2);
        \path (n2) edge[multiarc,out=0,in=180] (n3);
        \path (n2) edge[multiarc,out=0,in=180] (n4);
        \path (n2) edge[arc] (n5);
        \path (n3) edge[multiarc,out=0,in=180] (n6);
        \path (n4) edge[multiarc,out=0,in=180] (n6);

        % Init/Terminal vertices
        \path (i1.center) edge[arc] (n1);
        \path (n5) edge[arc] (t5.center);
        \path (n6) edge[arc] (t6.center);
      \end{tikzpicture} &
 % A not-independent F Graph
      %\tikzstyle{arc}=[->,draw, thick, color=gray]
      \begin{tikzpicture}
      %\pgftext{A F-Graph violating Indpendence Property}

        % Vertices, placed relatively
	\node[vertex] (n1) {1};
        \node[vertex,right=\HSpace of n1] (n2) {2};
        \node[vertex,right=\HSpace of n2,yshift=-2.2em] (n4) {4};
        \node[vertex,above=\VSpace of n4] (n3) {3};
        \node[vertex,right=\HSpace of n3,yshift=-2.2em] (n6) {6};

	% Init and terminal states
        \node[left=\VInitTermSpace of n1] (i1) {};
        \node[right=\VInitTermSpace of n6] (t6) {};

        % Arcs
        \path (n1) edge[arc] (n2);
        \path (n2) edge[multiarc,out=0,in=180] (n3);
        \path (n2) edge[multiarc,out=0,in=180] (n4);
        \path (n3) edge[arc] (n6);
        \path (n4) edge[arc] (n6);

        % Init/Terminal vertices
        \path (i1.center) edge[arc] (n1);
        \path (n6) edge[arc] (t6.center);

      \end{tikzpicture}
\\
      General hypergraph& Acyclic F-graph failing the independence property\\[1em]
      \begin{tikzpicture}
      %\pgftext{A F-Graph}
        % Vertices, placed relatively
        \node[vertex] (n1) {1};
        \node[vertex,right=\HSpace of n1] (n3) {3};
        \node[vertex,above=\VSpace of n3] (n2) {2};
        \node[vertex,below=\VSpace of n3] (n4) {4};
        \node[vertex,right=\HSpace of n2] (n6) {6};
        \node[vertex,above=\VSpace of n6] (n5) {5};
        \node[vertex,below=\VSpace of n6] (n7) {7};

        % Init and terminal states
        \node[left=\VInitTermSpace of n1] (i1) {};
        \node[right=\VInitTermSpace of n3] (t3) {};
        \node[right=\VInitTermSpace of n4] (t4) {};
        \node[right=\VInitTermSpace of n5] (t5) {};
        \node[right=\VInitTermSpace of n6] (t6) {};
        \node[right=\VInitTermSpace of n7] (t7) {};

        % Arcs
        \path (n1) edge[harc] (n2);
        \path (n1) edge[harc] (n3);
        \path (n1) edge[harc] (n4);
        \path (n2) edge[hmultiarc,out=0,in=200] (n5);
        \path (n2) edge[hmultiarc,out=0,in=180] (n6);
        \path (n2) edge[hmultiarc,out=0,in=160] (n7);
        \path (n4) edge[harc] (n7);

        % Init/Terminal vertices
        \path (i1.center) edge[harc] (n1);
        \path (n3) edge[harc] (t3.center);
        \path (n4) edge[harc] (t4.center);
        \path (n5) edge[harc] (t5.center);
        \path (n6) edge[harc] (t6.center);
        \path (n7) edge[harc] (t7.center);
	
      \end{tikzpicture}&
      %All F Paths
      \tikzstyle{arc}=[->,dotted, very thick, color=gray]
      \tikzstyle{arc1}=[->,thick, color=gray]
      \tikzstyle{arc2}=[->,dashed, color=gray]
\scalebox{0.6}{
\begin{tikzpicture}[scale=.5]
% First path
  % Vertices, placed relatively
  \node[vertex] (p1n1) {1};
  \node[vertex,right=\HSpace of p1n1] (p1n2) {2};
  \node[vertex,right=\HSpace of p1n2] (p1n6) {6};
  \node[vertex,above=\VSpace of p1n6] (p1n5) {5};
  \node[vertex,below=\VSpace of p1n6] (p1n7) {7};
  % Arcs
  \path (p1n1) edge[harc] (p1n2);
  \path (p1n2) edge[hmultiarc,out=0,in=200] (p1n5);
  \path (p1n2) edge[hmultiarc,out=0,in=180] (p1n6);
  \path (p1n2) edge[hmultiarc,out=0,in=160] (p1n7);

  % Init and terminal states
  \node[left=\VInitTermSpace of p1n1] (p1i1) {};
  \node[right=\VInitTermSpace of p1n5] (p1t5) {};
  \node[right=\VInitTermSpace of p1n6] (p1t6) {};

  % Init/Terminal vertices
  \path (p1i1.center) edge[harc] (p1n1);
  \path (p1n5) edge[harc] (p1t5.center);
  \path (p1n6) edge[harc] (p1t6.center);

% Second path
  % Vertices, placed relatively
  \node[vertex] at (0,-3) (p2n1) {1};
  \node[vertex,right=\HSpace of p2n1] (p2n3) {3};
  % Arcs
  \path (p2n1) edge[harc] (p2n3);

  % Init and terminal states
  \node[left=\VInitTermSpace of p2n1] (p2i1) {};
  \node[right=\VInitTermSpace of p2n3] (p2t3) {};

  % Init/Terminal vertices
  \path (p2i1.center) edge[harc] (p2n1);
  \path (p2n3) edge[harc] (p2t3.center);

% Fourth path
  % Vertices, placed relatively
  \node[vertex] at (0,-6) (p4n1) {1};
  \node[vertex,right=\HSpace of p4n1] (p4n4) {4};
  \node[vertex,right=\HSpace of p4n4] (p4n7) {7};
  % Arcs
  \path (p4n1) edge[harc] (p4n4);
  \path (p4n4) edge[harc] (p4n7);

  % Init and terminal states
  \node[left=\VInitTermSpace of p4n1] (p4i1) {};

  % Init/Terminal vertices
  \path (p4i1.center) edge[harc] (p4n1);

% Fifth path
  % Vertices, placed relatively
  \node[vertex] at (0,-9) (p5n1) {1};
  \node[vertex,right=\HSpace of p5n1] (p5n4) {4};
  % Arcs
  \path (p5n1) edge[harc] (p5n4);

  % Init and terminal states
  \node[left=\VInitTermSpace of p5n1] (p5i1) {};
  \node[right=\VInitTermSpace of p5n4] (p5t4) {};

  % Init/Terminal vertices
  \path (p5i1.center) edge[harc] (p5n1);
  \path (p5n4) edge[harc] (p5t4.center);

\end{tikzpicture}}\\
      Typical acyclic and independent F-graph & Associated set of F-paths
      \end{tabular}}
    \end{center}
    \caption{Illustration of F-Graphs, F-Paths and Independence property. Straight lines indicate classic arcs, and bent lines indicate hyperarcs.}
  \end{figure}
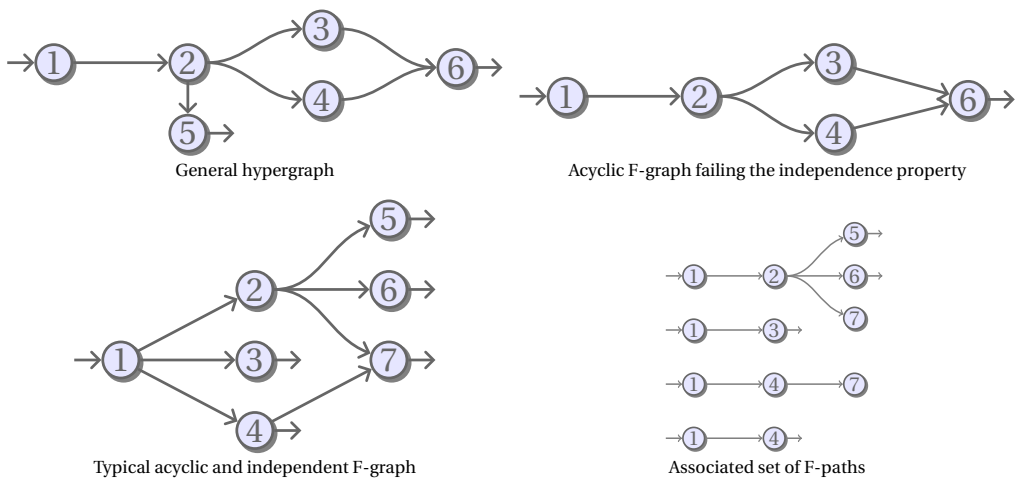

  Let us first remind that a directed hypergraph generalizes the notion of directed graph by allowing any number of vertices as origin(\Def{tail}) and destination (\Def{head}) for each (hyper)-arcs.
  We will be focusing here on Forward-Hypergraphs, or \Def{F-graphs}, which restrict the tail of their arcs to a single vertex.

    Formally, let $V$ be a set of vertices, an \Def{F-arc} $e=(t(e)\to {\bf h}(e)) \in V\times \mathcal{P}(V)$,
    connects a single tail vertex $t(e)\in V$ to an ordered list of vertices ${\bf h}(e)\subseteq V$.
    An \Def{F-graph} $\HG=(V,E)$ is characterized by a set of vertices $V$  and a set of F-arcs $E$.
    Denote by $\Children{n}$ the children of a node in a tree, then an \Def{F-path} of $\HG=(V,E)$ is a
    tree $\HT=(V'\subseteq V,E')$ such that, for any node $n\in V'$, $(v_n \to \Children{n})\in E$. For the sake of simplicity,
    we may omit the implicit $V'$ and identify an F-path to its set of edges $E'$.

    An \Def{F-derivation} from
    a vertex $s\in V$ can be recursively defined as either $\langle s, \varnothing\rangle$ if $(s\to\varnothing) \in E$,
    or $\langle s, D_1\,\ldots\,D_{|{\bf t}|}\rangle$ if $(s\to{\bf t}) \in E$, ${\bf t} = \{t_1,t_2,\ldots,t_{|t|}\}$, and each $D_i$ is an F-derivation starting from $t_i$.
    An F-graph is \Def{acyclic} if and only if any vertex $s\in V$ is present only once (as a root) in any derivations starting from $s$.
    Moreover it is \Def{independent} if and only if any vertex $s\in V$ is reached at most once in any derivation, regardless of its root.

    A \Def{weighted F-graph} is a triplet $(V,E,\W)$ such that $(V,E)$ is an F-graph and $\W: E \to \mathbb{R}^+$ is a weight function that associates
    a weight to each F-arc. Finally, an {\bf oriented F-graph} is a quadruplet $(\Init,V,E,\W)$ such that $(V,E,\W)$ is a weighted independent F-graph,
    and $\Init\in V$ is a distinguished initial vertex.
  \medskip

  {\noindent{\bf Remark~1:} Notice that our definition of F-arcs and F-paths implicitly defines {\bf terminal vertices}, since any leaf $l$ in a F-path has no child
  and our definition of F-paths therefore requires $l \to \varnothing$ to be an F-arc of $\HG$.}
  \medskip

  {\noindent{\bf Remark~2:} Under the independence property, the derivations starting from any node $s\in V$ are trees, and are therefore in bijection with F-paths originating
  from the same vertex. }

  \section{Generic problems and algorithms for F-paths in F-graphs}\label{sec:generic}
  \label{sec:probalgo}
  In the following, terminal cases will very seldom appear explicitly, but will rather be captured by the limit cases of products
  $\prod_{u\in \varnothing} f(u) = 1$ and sums $\sum_{u\in \varnothing} f(u) = 0$, $k\in \mathbb{R}$.
  \subsubsection{Generating and counting F-paths in oriented F-graphs~\cite{Wilf1977}}
  Let $\HG=(\Init,V,E,\W)$ be an oriented F-graph, we address the problem of generating the set $\FPath{\Init}$ of F-paths obtained starting from $\Init$.

  From the tree-like definition of F-paths and our remark on terminal vertices, we know that any F-path starting from a vertex $s$ can either be a leaf, provided that
  there exists an F-arc $s \to \varnothing$, or an internal node. In the latter case, any F-paths is composed of auxiliary paths, generated from the vertices in the head of
  some F-edge having $s$ as tail. Remark that our definition of F-paths requires each vertex from $V$ to appear at most once in any F-path, a fact that is ensured here by the acyclicity of $\HG$.
  Therefore we can recursively define the set of $\FPath{s}$ of F-paths starting from a root node $s$ as
  \begin{eqnarray}
    \FPath{s} & = & \left\{\begin{array}{cl}
      {\{(s,\varnothing)\}} & \text{If }(s,\varnothing)\in E\\
      \varnothing & \text{Otherwise}\end{array} \right\} \;\cup\;
      \bigcup_{(s\to {\bf t}) \in E}\left( \{s\}\times \prod_{u\in {\bf t}}\FPath{u}\right) ,\quad \forall s\in V.\label{eq:pathSets}
  \end{eqnarray}
  Since $E$ is a set, the candidate heads for a given tail $s$ are distinct and the unions in the above equations are disjoint. Furthermore,
  the products are Cartesian, so we can directly transpose the recurrence above over the cardinalities $\NumFPath{s}=|\FPath{s}|$ and obtain
  \begin{eqnarray}
    \NumFPath{s} & = & \sum_{(s\to{\bf t}) \in E} \prod_{u\in {\bf t}} \NumFPath{u},\quad \forall s\in V.\label{eq:nbPaths}
  \end{eqnarray}
  This immediately yields a $\Theta(|V|+|E|+\sum_{e\in E} |{\bf h}(e)|)/\Theta(|V|)$ time/memory dynamic programming algorithm for counting F-paths.

  \subsubsection{Minimal score F-path}
    Let us consider an \Def{additive scoring scheme} based on weights, and accordingly define the \Def{score} of an F-path $p$ to be
    $\Score{p} = \sum_{e\in E}\W(e)$. We address here the problem of finding an F-path $p_0$ having minimal score or more formally
    some $p_0 \in \FPath{\Init}$ such that $\forall p\in\FPath{\Init} ,\;p\neq p_0 \Rightarrow \Score{p}\ge \Score{p_0}$.
    From the independence of siblings and the strict additivity of the score, we know that the
    path minimization problem has optimal substructure, i. e. any optimal solution is composed of optimal solutions for its subproblems.
    Consequently, the \Def{minimal score} $\MinFPath{s}$ of a path starting from a root node $s\in V$ is given by
    \begin{eqnarray}
      \MinFPath{s} & = & \min_{e=(s\to{\bf t}) \in E} \left(\W(e)+\sum_{u\in {\bf t}}{\MinFPath{u}}\right),\quad \forall s\in V.\label{eq:minscore}
    \end{eqnarray}
    A classic backtrack procedure can then be used to reconstruct the F-path instance $\ArgMinFPath{s}$ starting from $s\in V$  and having minimal score.
    Alternatively, the previous recurrence can be modified as follows
    \begin{eqnarray}
      \ArgMinFPath{s} & = & \argmin_{\substack{
      p' = \bigcup_{s' \in {\bf t}} \ArgMinFPath{s'}\\
      \text{s.t. } (s\to{\bf t}) \in E} } \Score{\left\{(s \to {\bf t})\right\} \cup p'},\quad \forall s\in V,
    \end{eqnarray}
    giving a $\Theta(|V|+|E|+\sum_{e\in E} |{\bf h}(e)|)$/$\Theta(|V|)$ time/memory DP algorithm for the minimal weighted F-path.

  \subsubsection{Weighted count and weighted random generation~\cite{Denise2010}}
  Let us \Def{extend multiplicatively on paths} our weight function, defining the \Def{weight of any F-path} $p$ to be $\W(p) = \prod_{e\in p} \W(e)$.
  Then a small modification of Equation~\ref{eq:nbPaths} gives a recurrence for computing the cumulated weight, or \Def{weighted count} $\WNumFPath{s}$ of F-paths starting from
  a given vertex $s$:
  \begin{eqnarray}
    \WNumFPath{s} & = &  \sum_{p'\in\FPath{s}}\W(p') = \sum_{e=(s\to{\bf h}(e)) \in E} \W(e)\cdot \prod_{s'\in {\bf h}(e)} \WNumFPath{s'},\quad \forall s\in V\label{eq:weightedCount}
  \end{eqnarray}

  Provided that the weights are positive, this defines a \Def{weighted probability distribution} over F-paths, which assigns
  to each path $p\in\FPath{\Init}$ a probability
  \begin{equation}\Prob{p\;|\;\W} = \frac{\W(p)}{\sum_{p'\in\FPath{\Init}}\W(p')} \equiv  \frac{\W(p)}{\WNumFPath{\Init}}.\label{eq:WeightedPD}\end{equation}
  %reminding that $\WNumFPath{\Init}$ is the cumulated weight of all paths starting from $\Init$.

  From the precomputed values $\WNumFPath{s}$, one can perform a {\bf weighted random generation} to draw at
  random a set of $k$ F-paths from $\Init$ according to a weighted distribution.
  Starting from any vertex $s$, the algorithm chooses at each step an F-arc $e=(s \to {\bf h}(e))$ with probability
  $$p_{s,e} = \frac{\W(e)\cdot\prod_{s'\in {\bf h}(e)}\WNumFPath{s'}}{\WNumFPath{s}},$$
  and proceeds to the recursive generation of auxiliary paths from each vertex in ${\bf h}(e)$.
  A simple induction argument shows that any F-path is then generated with respect to the probability distribution of Equation~\ref{eq:WeightedPD}.
   The weighted count recurrence is computed by a $\Theta(|V|+|E|+\sum_{e\in E} |{\bf h}(e)|)$/$\Theta(|V|)$ time/memory algorithm,
   and each path $p$ is generated in $\Theta(|p|+\sum_{e\in p} |{\bf h}(e)|)$/$\Theta(|p|)$ time/memory.
  \medskip

  \noindent {\bf Remark~3:} This worst-case complexity can be improved using additional information on the structure of the F-graph. For instance,
  when both the height and maximal degree of a vertex are bounded by some constant $n$, Boustrophedon search~\cite{Flajolet1994,Ponty2008} can be used
  to decrease the worst-case complexity of each generation from $\mathcal{\Theta}(n^2)$ to $\mathcal{O}(n\log n)$.

  \subsubsection{Arc traversal probabilities}
  Using the same probability distribution, a natural problem is to compute the probability $p_e$ of an F-arc $e\in E$ being in a random F-path.
  To that purpose one can use the classic \emph{inside/outside} algorithm, which can be rephrased as an F-graphs traversal.

  Let us first point out that the probability $p_e$ is related to the cumulated weight of all F-paths featuring an edge $e=(t(e) \to {\bf h}(e))$ through
  \begin{equation}
  p_e = \frac{\sum_{\substack{p\in\FPath{\Init}\\\text{s.t. } e\in p} } \W(p) }{\sum_{p'\in\FPath{\Init}}\W(p')}
  \equiv \frac{\sum_{\substack{p\in\FPath{\Init}\\\text{s.t. } e\in p} } \W(p) }{\WNumFPath{\Init}}.
  \end{equation}
  From the independence of $\HG$, we know that each vertex appears at most once in any given F-path, and consequently any F-path traversing $e$
  can therefore be {\bf unambiguously} decomposed into: i) An \Def{$\bf e$-outside tree}, i.e. a derivation from $\Init$ whose leaves are either terminal or $t(e)$,
  and which features exactly one occurrence of $t(e)$; ii) A \Def{support edge} $e=(t(e) \to {\bf h}(e))$; iii) An \Def{$\bf e$-inside tree}, i.e. a set of F-paths issued from ${\bf h}(e)$.

  The unambiguity of the decomposition, along with the independence of i) and iii), translates into
    \begin{equation}\sum_{\substack{p\in\FPath{\Init}\\\text{s.t. } e\in p} } \W(p) = \OutsideWeights{t(e)}\cdot \EdgeWeight{e}\cdot \prod_{s'\in {\bf h}(e)} \WNumFPath{s'}\end{equation}
  where $\OutsideWeights{s}$ is the cumulated weight of all outside trees leaving $s\in V$ underived.
  Finally it can be shown that the cumulated weight $\OutsideWeights{s}$ over all $e$-outside trees obey the following simple recurrence
    \begin{equation}\OutsideWeights{s} = \mathbf{1}_{s=q_0} + \sum_{\substack{e' \in E\\ \text{s. t. } s\in {\bf h}(e')}} \W(e')\cdot\OutsideWeights{t(e')}\cdot\prod_{\substack{s' \in {\bf h}(e')\\s'\neq s}} \WNumFPath{s'},\quad \forall s\in V\end{equation}
  which can computed in  $\BigO{|V|+|E|+\sum_{e\in E} |{\bf h}(e)|^2}$/$\Theta(|V|)$ time/memory. The probability of traversing $p_e$ in a random
  F-path can finally be computed through the formula
  \begin{equation} p_e = \frac{\OutsideWeights{t(e)}\cdot \prod_{s'\in {\bf h}(e)} \WNumFPath{s'}}{\WNumFPath{\Init}},\quad \forall e\in E.\end{equation}

  \section{F-graphs reformulation of (Pseudoknotted) RNA conformation spaces}\label{sec:conformations}
    From the previous section, we know that very simple algorithms exist for weighted optimization and enumeration problems over the F-paths of an F-graph.
    Let us now consider MFE folding-related problems over an arbitrary \Def{conformation space} $\ConfSpace$ for a sequence $\RNA$, under an energy model $E: \ConfSpace \to \mathbb{R}$
    and assume that there exists: {\bf C1.} An F-graph $\HG$ whose F-paths $\FPath{}$ are in bijection with the conformation space $\ConfSpace$;
    {\bf C2.} A weight function $\W$ such that the (additive) score of any F-path coincides with the energy of its corresponding conformation.

    Under such conditions, it can be remarked that the {\bf minimal score} algorithm (Equation~\ref{eq:minscore}) exactly computes the {\bf Minimal Free-Energy} $MFE=\min_{s\in \ConfSpace} E_s$.
    Furthermore, the {\bf Weighted Count} (Equation~\ref{eq:weightedCount}), applied to a weight function $\W'(e)=e^{-\W(e)/RT}$, computes
    the {\bf Partition Function} $\mathcal{Z}=\sum_{s\in \ConfSpace} e^{-E_s/RT}$. Other quantities of interest for RNA folding can also be derived, as summarized
    in Tables~\ref{tab:genericUnafold} and \ref{tab:recapAlgo}.

  \subsection{Foreword: Shortening correctness proofs through generating functions}

    Our main challenge is to find an hypergraph/weight such that the energy function can be expressed in an additive fashion.
    Focusing first on Condition {\bf C1}, one remarks that finding a function $\psi:\FPath{}\to \ConfSpace$ which maps F-Paths to elements of the conformation space is not challenging, as it essentially amounts to figuring out which derivation
    creates which base-pairs. Condition {\bf C1} is then traditionally broken into two parts: an \Def{unambiguity} condition which requires distinct elements in $\FPath{}$ to give rise
    to distinct elements within $\ConfSpace$, i.e. $\psi$ should be injective; a \Def{completeness} condition which requires each element in $S$ to have at least one pre-image,
    i.e. $\psi$ should be surjective.

    Since these notions are intimately related to the semantics associated with the F-paths, they cannot be tackled in an automated way at the hypergraph
    level\footnote{Algebraic Dynamic Programming partially addresses this issue, and the interested reader is referred to an early contribution by
    Reeder~\emph{et al}~\cite{Reeder2005a}.}. Therefore correctness proofs will usually require user-assigned semantics coupled
    with custom arguments, a task that may become challenging and/or tedious for complex decompositions. In order to simplify the validation and therefore the design
    of new conformation spaces, we propose a simplified proof technique based on generating functions.

    Indeed, instead of specializing the hypergraph for each and every input sequence, one can delegate to the weight function the responsibility
    of weeding out conformations, e.g. by assigning them $+\infty$ energetic contributions within MFE folding. Therefore each
    class of conformations can be seen as a family of conformation space $\{\ConfSpace_n\}_{n\ge0}$ (secondary structures, simple type pseudoknots\ldots), to which one associates
    a family of hypergraphs $\{\HG_n\}_{n\ge0}$, a \Def{decomposition}, both indexed by the length $n$ of the sequence.

    Let us remind that generating functions are formal power series that can be used to store various information. For instance the counting generating function
    for the conformation space family $\ConfFamily$ can be defined as $S_{\ConfFamily}(z) = \sum_{n\ge0}|\ConfSpace_n|\cdot z^{n}$ where $z$ is a formal complex variable devoid
    of intuitive meaning. Furthermore let $\FPath{n}$ be the set of F-Paths associated with $\HG_n$, then the counting generating function of the decomposition
     can be defined as $S_{\HG}(z) = \sum_{n\ge0}|\FPath{n}|\cdot z^{n}$. Then the formal identity $S_{\ConfFamily}(z) = S_{\HG}(z)$ implies that $|\ConfSpace_n|=|\FPath{n}|, \forall n\ge0$.
     It follows from basic set theory that unambiguity/injectivity (resp. completeness/surjectivity) of $\psi$, in addition to the identity of generating functions, is in itself sufficient to prove the bijectivity of $\psi$. Since reference generating functions are now available for many conformation space families~\cite{Saule2010}, this practically halves the burden of designing a proof.
  \label{sec:fgraformul}
  \begin{figure}[t]
  \centering \includegraphics[width=\textwidth]{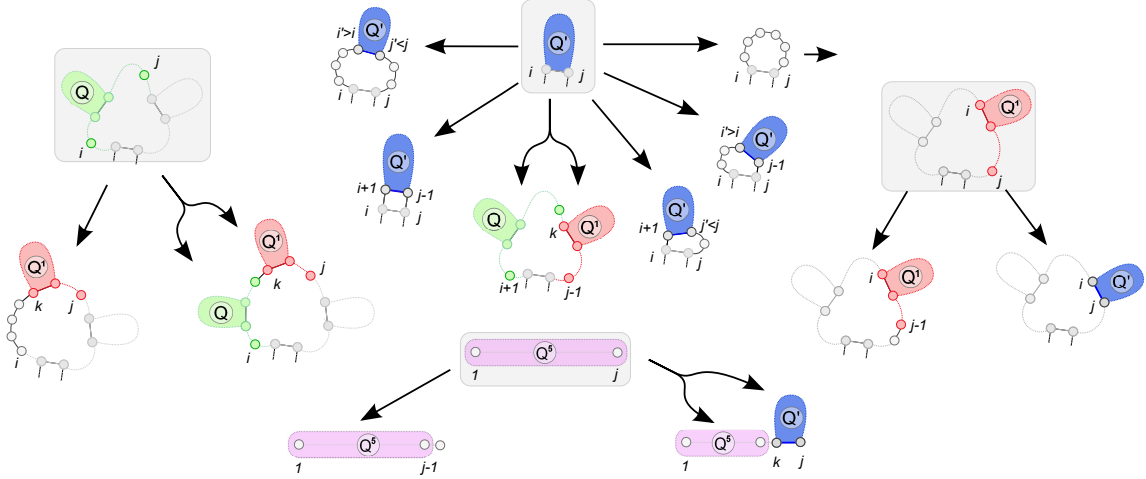}\\
  \caption{Simplification of the \Soft{Unafold}~\cite{Markham2008a} decomposition of the secondary structures space.
  Framed states indicate origins of (hyper)arcs. \label{fig:unafold}}
  \end{figure}
    \subsection{RNA secondary structures}

     Let us first illustrate our approach on RNA secondary structures, for which \Soft{Unafold}~\cite{Markham2008a} -- the
     successor of \Soft{MFold}~\cite{Zuker1981} -- offers
     an unambiguous scheme.  Compared to the original decomposition presented in Markham's thesis~\cite{Markham2006}, the one described in Figure~\ref{fig:unafold} is simplified to ignore dangles.
     %\Yann{Reminder: If there is an extended version, we should add discuss the opportunity of accounting for dangles within partition function-based algorithms.}

   \subsubsection{Proving unambiguity.}
  \begin{itemize}
    \item Let us remark that both $Q^5$ and $Q^1$ either leave their last base $j$ unpaired (Left), or pairs it to $i$ (Right). Furthermore these two cases are mutually exclusive. Finally $Q^1$ generates exactly one helix.
    \item $Q$ always makes at least one call to $Q^1$ and therefore creates at least one helix. Therefore, it either creates exactly one helix (Left case) or more (Right case), and these two cases are mutually exclusive.
        \item $Q'$ distinguishes different types of loops. Let $m_5$, $m_3$ be the numbers of unpaired bases on the $5'$ strand, $3'$ strand, and $h$ be the number of helices starting from case $Q'$, one can label each of the cases and observe that they are mutually non-overlapping. Namely from left to right, we get the following $(m_5,m_3,h)$ triplets: Interior loop $(>0,>0,1)$, stacking pair $(0,0,1)$, multiloop $(\ge 0,\ge 0,>1)$, bulges $5'$ $(>0,0,1)$ and $3'$ $(0,>0,1)$, and hairpin loop $(>0,>0,0)$.
  \end{itemize}

  \subsubsection{Deriving completeness.}
  From previous work by Waterman~\cite{Waterman1978}, we know that the generating function of secondary structures with at least one unpaired base
  between paired bases ($\theta=1$) is
  \begin{equation} S(z) = \frac{1-z+z^2-\sqrt{1-2z-z^2-2z^3+z^4}}{2z^2}.\end{equation}

  Following the general principle of the so-called DSV methodology (See Lorenz~\emph{et al}~\cite{Lorenz2008} for a presentation in a similar context),
  the \Soft{Unafold} decomposition can be translated into a system of algebraic equations.
  Namely, one simply replaces any occurrence of $k$ unpaired base with $z^k$, each basepair with $z^2$, and any vertex with its associated generating function.
  Let $Q^5(z)$, $Q(z)$, $Q'(z)$ and $Q^1(z)$ be the generating functions
  counting the F-paths generated from $Q^5$, $Q$, $Q'$ and $Q^1$ respectively:
  \begin{align*}
    Q^5(z)  = &Q^5(z)\cdot z+ Q^5(z)\cdot Q'(z) \quad\quad  Q(z)  = \Seq(z)\cdot Q^1(z) + Q(z)\cdot Q^1(z) \quad\quad  Q^1(z)  = &z\cdot Q^1(z) + Q'(z)\\
    Q'(z)  = &z^2\cdot \Seq^+(z)\cdot Q'(z)\cdot\Seq^+(z)
    +z^2\cdot Q'(z)
    +z^2\cdot Q(z)\cdot Q'(z)\\
    & + z^2\cdot Q'(z)\cdot\Seq^+(z)+z^2\cdot \Seq^+(z)\cdot Q^1(z)+\Seq^+(z)\\
    \Seq^+(z)  = &z\cdot\Seq(z) \quad\quad  \Seq(z)  = z\cdot\Seq(z) + 1.
  \end{align*}
  Solving the system yields $Q^5(z) =  S(z)$ which, in conjunction with the unambiguity of the decomposition, proves its completeness.
      \begin{table}[t]
       {\centering\begin{tabular}{|l|l|cccc|}\hline
         Application         &  Algorithm      & Weight fun.& Time & Memory & Ref.\\ \hline \hline
         A -- Energy minimization &  Minimal weight & $\W_{\Turner}$  & $\BigO{n^3}$ & $\BigO{n^2}$ & \cite{Zuker1981}\\
         B -- Partition function  &  Weighted count & $e^{\frac{-\W_{\Turner}}{RT}}$  & $\BigO{n^3}$ & $\BigO{n^2}$ & \cite{McCaskill1990}\\
         C -- Base-pairing probabilities  &  Arc-traversal prob. & $e^{\frac{-\W_{\Turner}}{RT}}$  & $\BigO{n^3}$ & $\BigO{n^2}$ & \cite{McCaskill1990}\\
         D -- Statistical sampling ($k$-samples) & Weighted random gen. & $e^{\frac{-\W_{\Turner}}{RT}}$ & $\BigO{n^3+k\cdot n\log n}$ & $\BigO{n^2}$ & \cite{Ding2003,Ponty2008}\\
         E -- Moments of energy (Mean, Var.) &  Moments extraction & $e^{\frac{-\W_{\Turner}}{RT}}$ & $\BigO{n^3}$ & $\BigO{n^2}$ & \cite{Miklos2005}\\\hline
         F -- $m$-th moment of additive features & Moments extraction & $e^{\frac{-\W_{\Turner}}{RT}}$ & $\BigO{m^3\cdot n^3}$ & $\BigO{m\cdot n^2}$ & -- \\
         G -- Correlations of additive features & Moments extraction & $e^{\frac{-\W_{\Turner}}{RT}}$ & $\BigO{n^3}$ & $\BigO{n^2}$ & -- \\ \hline
       \end{tabular}\\}
       \caption{Reformulations of secondary structure applications as F-graphs problems and associated complexities.
       \label{tab:genericUnafold}}
      \end{table}     \subsubsection{Applicability of generic algorithms.} Let us show that $\HG$ fulfills the prerequisites of our algorithms.
     First it is easily verified that $\HG$ is an F-graph.
     Associating a region $[i,j]$ (resp. $[1,j]$) with each vertex $q^1_{i,j}$, $q_{i,j}$ and $q'_{i,j}$  (resp. $q^5_{j}$), one easily verifies
     that for any F-arc $e\in E$ the width of any region in the head ${\bf h}(e)$ is strictly smaller than that
     of the tail $t(e)$, and the {\bf acyclicity} of $\HG$ directly follows.
     Furthermore, any two vertices in the head ${\bf h}(e)$ have non-overlapping associated regions.
     Consequently $\HG$ is {\bf independent}, and a direct application of our generic algorithms gives a set of algorithms summarized in Table~\ref{tab:genericUnafold}.
      This gives a family of efficient $\BigO{n^3}$ algorithms for assessing RNA secondary structure properties at the Boltzmann equilibrium.

  \begin{figure}[!h]
    {\centering \includegraphics[width=.8\textwidth]{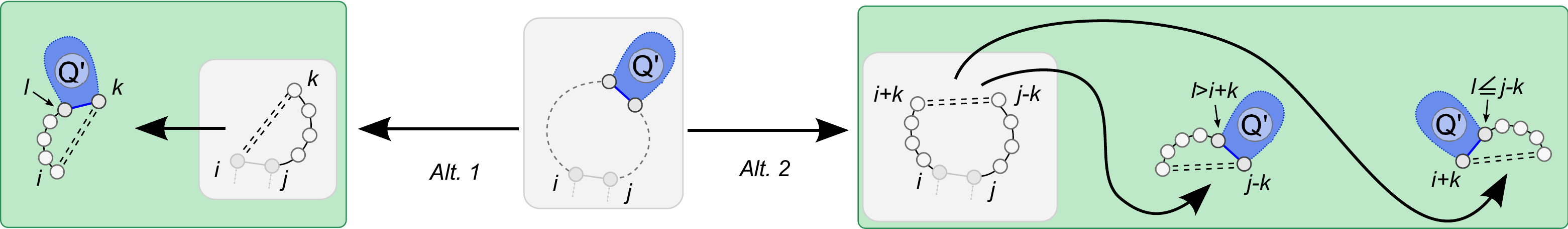} \\}
    \caption{Alternative exhaustive strategies for interior loops.\label{fig:interior}}
  \end{figure}
  \noindent {\bf Remark~4:} In interior loops, the set of F-arcs generated for the $Q'$ case has apparent cardinality in $\BigO{n^4}$.
  This can be brought back to $\BigO{n^3}$ by enforcing constraints on the energy function.
  Traditionally, the accepted practice is to bound the interior loop \emph{size} $(j'-j)+(i'-i)$ from above by a predefined constant $K\approx 30$.
  Exhaustive $\BigO{n^3}$ decompositions can also be proposed (Figure~\ref{fig:interior}) by decomposing the internal loop into
  additively-contributing regions. A first option may generate independently the left and right unpaired regions (Figure~\ref{fig:interior}, Left),
  while an alternative may decompose internal loops into a symmetric loop followed by a fully asymmetric one (Figure~\ref{fig:interior}, Right).

%%%%%%%%%%%%%%%%%%% BEGIN CEDRIC PART %%%%%%%%%%%%%%%%%%%%%%%%%%%%%

\input{unambiguousCedric}

  \section{Extending the framework: Extraction of moments and exact correlations}\label{sec:moments}
  A last application addresses the extraction of statistical measures for \Def{additive features}. Let us first define a \Def{feature}
  as a function $\FeatureFun: E \to \mathbb{R}^+$ extended additively over F-paths such that $\Feature{p} = \sum_{e\in p} \Feature{e}$. One may then want
  to characterize the distribution of a random variable $X=\Feature{p}$, for $p\in\FPaths$ a random F-path drawn according to the weighted distribution.
  As it is not necessarily feasible to determine the exact distribution of $X$, one can examine statistical measures such as its
  $$\text{Mean } \Mean{X} = \Expect{X}  \quad\text{ and }\quad \text{Variance }\Var{X} = \Expect{X^2}-\Mean{X}^2,$$
  e.g. from which the distribution is fully determined in the case of Gaussian distributions. Even when the distribution is not normal,
  it can still be characterized by a list of measures called \Def{moments} of $X$, the $m$-th moment being defined
  as $\Expect{X^m} = \sum_{p \in \FPaths} \FeatureFun(p)^{m}\cdot\W(p)/\WNumFPath{s}$.

  Moreover in the presence of  multiple features $(X_1:=\FeatureFun_1(p),\ldots,X_k:=\FeatureFun_k(p))$, similar measures can be used to estimate
  their level of dependency. One such measure is the \Def{Pearson product-moment correlation coefficient} $\rho_{X_1,X_2}$ defined for two
  random variables as
  $$ \rho_{X_1,X_2} = \frac{\Cov{X_1,X_2}}{\sqrt{\Var{X_1}\cdot\Var{X_2}}} = \frac{\Expect{X_1\cdot X_2}-\Expect{X_1}\cdot\Expect{X_2}}{\sqrt{\Var{X_1}\cdot\Var{X_2}}}$$

\begin{table}[t]
       {\centering\begin{tabular}{|l|c|cccc|}\hline
         Application         &  Algorithm      & Weight fun.& Time & Memory & Ref.\\ \hline
    \multicolumn{6}{c}{Simple type pseudoknots (Akutsu\&Uemura)}\\ \hline
         A -- Energy minimization &  Minimal weight & $\W_{\bp}$  & $\BigO{n^4}$ & $\BigO{n^4}$ & \cite{Akutsu2000}\\
         B -- Partition function  &  Weighted count & $e^{\frac{-\W_{\bp}}{RT}}$  & $\BigO{n^4}$ & $\BigO{n^4}$ & \cite{Cao2006,Cao2009} in $\Theta(n^6)$\\
         C -- Base-pairing probabilities  &  Arc-traversal prob. & $e^{\frac{-\W_{\bp}}{RT}}$  & $\BigO{n^4}$ & $\BigO{n^4}$ & --\\
         D -- Statistical sampling ($k$-samples) & Weighted rand. gen. & $e^{\frac{-\W_{\bp}}{RT}}$ & $\BigO{n^4+k\cdot n\log n}$ & $\BigO{n^4}$ & --\\
         E -- Moments of energy (Mean, Var.) &  Moments extraction & $e^{\frac{-\W_{\bp}}{RT}}$ & $\BigO{n^4}$ & $\BigO{n^4}$ & --\\
         F -- $m$-th moment of additive features & Moments extraction & $e^{\frac{-\W_{\bp}}{RT}}$ & $\BigO{m^3\cdot n^4}$ & $\BigO{m\cdot n^4}$ & -- \\ \hline
    \multicolumn{6}{c}{Fully recursive Kissing Hairpins}\\ \hline
         A -- Energy minimization &  Minimal weight & $\W_{\Turner}$  & $\BigO{n^5}$ & $\BigO{n^4}$ & \cite{Chen2009}\\
         B -- Partition function  &  Weighted count & $e^{\frac{-\W_{\Turner}}{RT}}$  & $\BigO{n^5}$ & $\BigO{n^4}$ & --\\
         C -- Base-pairing probabilities  &  Arc-traversal prob. & $e^{\frac{-\W_{\Turner}}{RT}}$  & $\BigO{n^5}$ & $\BigO{n^4}$ & --\\
         D -- Statistical sampling ($k$-samples) & Weighted rand. gen. & $e^{\frac{-\W_{\Turner}}{RT}}$ & $\BigO{n^5+k\cdot n\log n}$ & $\BigO{n^4}$ & --\\
         E -- Moments of energy (Mean, Var.) &  Moments extraction & $e^{\frac{-\W_{\Turner}}{RT}}$ & $\BigO{n^5}$ & $\BigO{n^4}$ & --\\
         F -- $m$-th moment of additive features & Moments extraction & $e^{\frac{-\W_{\Turner}}{RT}}$ & $\BigO{m^3\cdot n^5}$ & $\BigO{m\cdot n^4}$ & -- \\ \hline
       \end{tabular}\\}
       \caption{Summary of ensemble based algorithms on simple pseudoknots and kissing hairpins.
       $\W_{\bp}$ stands for the simple Nussinov-Jacobson energy model, and $\W_{\Turner}$ for a Turner-like model based on loops contributions.
       \label{tab:recapAlgo}}
      \end{table}

  The correlation above involves the expectation of a product of two random variables which is an instance of a \Def{generalized moment}, defined for the set of F-paths starting from $s\in V$ as
  \begin{equation} \Expect{X_{1}^{m_1}\cdots X_k^{m_k}\;|\;s} = \sum_{p \in \FPath{s}} \frac{\W(p)}{\WNumFPath{s}}\prod_{i=1}^{k}\FeatureFun_i(p)^{m_i}. \label{eq:moments}\end{equation}
  Extracting such moments can be quite useful, allowing one to get access to average properties of structures (\#Hairpins, \#Occurrences of pseudoknots\ldots) and
  their correlations within a weighted ensemble. For instance, Miklos~\emph{et al}~\cite{Miklos2005} proposed an $\mathcal{O}(m^2\cdot n^3)$ algorithm for computing the
  $m$-th moment of the Energy distribution for secondary structure in order to compare the
  distribution of free-energy in non-coding RNAs and random sequences. We are going to show how these generalized moments can be extracted
  directly through a generalization of the weighted count algorithm.

  \begin{theorem}\label{th:moments}
    Let ${\boldmath{\alpha}}:=(\alpha_1,\cdots,\alpha_k)$ be a vector of additive features and ${\bf m}:=(m_1,\cdots,m_k)$ be a $k$-tuple of natural integers.
    Then the pseudo-moment $\Moment{{\bf m}}_{s} := \Expect{X_{1}^{m_1}\cdots X_k^{m_k}\;|\;s} \cdot \WNumFPath{s}$ of $\boldmath{\alpha}$ in a weighted distribution
    can be recursively computed through
        \begin{align}
        \Moment{{\bf m}}_{s}  & =
        \sum_{e=(s\to {\bf t})}\W(e)\cdot
    \sum_{\substack{{\bf m'},\left({\bf m}''_1,\cdots,{\bf m}''_{|t|}\right)\\
    \text{s. t. } {\bf m}'+\sum_j{\bf m}''_j={\bf m}}}
    \prod_{i=1}^{k}{ m_i \choose m'_i,m''_{1,i}, \cdots ,m''_{|t|,i}}\cdot\FeatureFun_i(e)^{m'_i}\cdot \prod_{i=1}^{|t|}\Moment{{\bf m}''_i}_{t_i}
    \end{align}
    in  $\mathcal{O}\left((|E|+|V|)\cdot k\cdot t^{+}\cdot \prod_{i=1}^{k} m_i^{t^+ +1}\right)$ time complexity and $\Theta\left(|V|\cdot\prod_{i=1}^{k} m_i\right)$ memory
    where $t^+ = \max_{(s\to t)\in E}(|t|)$ is the maximal out-degree of an arc.
  \end{theorem}

  Adding this new generic algorithms automatically creates new applications for each an every conformation space as summarized in Figure~\ref{tab:recapAlgo}.
  This simultaneous extension -- for all conformational spaces -- of possible ensemble applications constitues in our opinion one of the main benefit of detaching the decomposition
  from its exploration.

%%%%%%%%%%%%%%%%%%%%%%%%%%%%%%%%%%%%%%%%%%%%%%%%%%%%%%%%%%%%%%%%%%%%%%%%%%%%%%%%%%%%%%%
\section{Conclusion and Perspectives}
\label{sec:conclu}

In this paper, we established the foundation of a combinatorial approach to the design of
algorithms for complex conformation spaces. We built on an hypergraph model introduced in the context of
RNA secondary structure by Finkelstein and Roytberg~\cite{Finkelstein1993}, which we extended in
several direction. First we formulated classic and novel generic algorithms on Forward-Hypergraphs for weighted ensembles,
allowing one to derive base-pairing probabilities, perform statistical sampling and extract moments of the distribution
of additive features. Then we showed how combinatorial arguments based on generating functions could be used to
simplify the proof of correctness for designed decompositions. We illustrated the full programme on classic secondary
structures, simple type pseudoknots and fully-recursive kissing hairpin pseudoknots for which we provided decompositions
that were proven to be unambiguous and complete with respect to previous work. The hypergraph formulation of the
decomposition, coupled with the generic algorithms, readily gave a family of novel algorithms for complex -- yet
relevant -- conformation spaces.

Let us mention some perspectives to our contribution. Firstly the principles and algorithms described here
could easily  be implemented as a general \emph{compiler} tools for F-Graphs algorithms.
Such a compiler could be coupled with helper tools expanding hypergraphs from succinct descriptions, such as
 context-free grammars (related to ADP~\cite{Giegerich2000}), or
M. Möhl's split types~\cite{Moehl2010}. More complex search space could also be modeled, such as those relying
on a more detailed representation of RNA structure (e.g. MCFold's NCMs~\cite{Parisien2008}), those capturing
RNA-RNA interactions~\cite{Alkan2005,Huang2010}, those offering simultaneous alignment and folding (Sankoff's algorithm~\cite{Sankoff1985})
or performing mutations on the sequence~\cite{Waldispuehl2008}.
Finally our hypergraph framework is not necessarily limited to polynomial algorithms,
and algorithmic developments could be proposed to address some of the current algorithmic issues
in RNA (inverse folding~\cite{Andronescu2004}, kinetics~\cite{Thachuk2010}) for which no exact polynomial
algorithms are currently known (or suspected). More generally it is our hope that, by simplifying and modularizing the process of developing
new -- algorithmically tractable -- conformation spaces, our contribution will help design better, more topologically-realistic\cite{Vernizzi2006,Lescoute2006a,Reidys2011}, energy and conformational
spaces to better understand and predict the structure(s) of RNA.

\section*{Acknowledgement}
The authors wish to express their gratitude to M. Roytberg for pointing out his work on hypergraphs as a unifying framework,
and to R. Backofen, M. Möhl and S. Will for fruitful discussions.
This research was supported by the Digiteo project ``RNAomics''.
YP was funded by an ANR grant MAGNUM (ANR 2010 BLAN 0204).
% ----------------------------------------------------------------
\bibliographystyle{plain}
\bibliography{biblio}

\end{document}

%% file: macros.tex
  \DeclareMathOperator*{\argmin}{arg\,min}
  \DeclareMathOperator{\varop}{Var}
  \DeclareMathOperator{\covop}{Cov}
  \DeclareMathOperator{\Seq}{Seq}

  \def \oc {c}

  \def \ox {f}
  \def \bx {{\bar f}}
  \def \oy {g}
  \def \by {{\bar g}}
  \def \oz {h}
  \def \bz {{\bar h}}

  \newcommand{\RNA}{\omega}

  \newcommand{\MFold}{{\sc MFold}\xspace}
  
  \newcommand{\SFold}{{\sc SFold}\xspace}
  \newcommand{\CentroidFold}{{\sc CentroidFold}\xspace}
  \newcommand{\ContraFold}{{\sc ContraFold}\xspace}
  \newcommand{\Vienna}{{\sc ViennaRNA}\xspace}
  \newcommand{\ParHead}[1]{{\bf\noindent #1}\xspace}
  \newcommand{\ConfSpace}{D}
  \newcommand{\ConfFamily}{\mathcal{D}}

  \newcommand{\HG}{\mathcal{H}}
  \newcommand{\HT}{\mathcal{T}}
  \newcommand{\Init}{v_0}
  
  \newcommand{\Children}[1]{{\bf c}_{#1}}
  \newcommand{\NumFPath}[1]{n_{#1}}
  \newcommand{\WNumFPath}[1]{w_{#1}}
  \newcommand{\MinFPath}[1]{m_{#1}}
  \newcommand{\ArgMinFPath}[1]{p^{\min}_{#1}}
  \newcommand{\FPath}[1]{\mathcal{P}_{#1}}
  \newcommand{\FPaths}{\mathcal{P}}
  
  \newcommand{\Score}[1]{\alpha\left(#1\right)}
  
  \newcommand{\EdgeWeight}[1]{\W(#1)}

  \newcommand{\FeatureFun}{\alpha}
  
  \newcommand{\Feature}[1]{\FeatureFun({#1})}
  \newcommand{\OutsideWeights}[1]{b_{#1}}
  
  \newcommand{\W}{\pi}
  \newcommand{\Turner}{\mathcal{T}}

  \newcommand{\Expect}[1]{\mathbb{E}[#1]}
  \newcommand{\Mean}[1]{\mu_{#1}}
  \newcommand{\Moment}[1]{c^{#1}}
  
  \newcommand{\Var}[1]{\varop_{#1}}
  \newcommand{\Cov}[1]{\covop_{#1}}

  \newcommand{\Soft}[1]{{\tt #1}\xspace}
  \newcommand{\bp}{bp}

\tikzstyle{arc}=[-angle 90,draw, line width=1.3pt, color=gray!80!black]
\tikzstyle{harc}=[arc, solid]
\tikzstyle{multiarc}=[arc,looseness=1.]
\tikzstyle{hmultiarc}=[multiarc,solid]
\tikzstyle{vertex}=[color=black!60, circle, draw=black!60,fill=blue!10,line width=1.3pt, inner sep=0pt,minimum size=6mm,font=\LARGE,circular drop shadow]
\tikzstyle{pvertex}=[vertex, fill=white]
\tikzstyle{dot}=[midway, draw,fill, color=black!60, circle, inner sep=0pt,minimum size=1.5mm]
\tikzstyle{tdot}=[dot,pos=0.4]
\tikzstyle{mdot}=[dot,pos=0.2]
\tikzstyle{hyp}=[->,draw, thick, color=gray!80!black,decoration={snake,segment length=7pt,amplitude=1.5pt},decorate]
\tikzstyle{arclabel}=[ pos=.4,circle, draw=none, fill=white, fill opacity=0.8,inner sep=0pt,minimum size=3mm]

\newcommand{\BigO}[1]{{O}(#1)}
\newcommand{\Def}[1]{{\bf #1}}

\newcommand{\Prob}[1]{\mathbb{P}(#1)}

%% file: unambiguousCedric.tex
\begin{comment}

\section{Unambiguous decompositions for RNA pseudoknotted conformations}

In this section, We will present the unambiguous decomposition of of two RNA pseudoknotted conformations corresponding to simple pseudoknots and kissing hairpins.
Firstly, we characterize the language of structures we want to predict and we show that the algorithm
predicts them effectively. Secondly, we show that our decompositions of pseudoknots are not ambiguous.
To show it, we use the following steps :

\begin{itemize}
\item{Encode the pseudoknotted structure by a sub-word of the Dyck language.}
\item{Build a bijection between this langage and an unambiguous context-free language.}
\item{Deduce the algebraic equation $F_1$ for the ordinary generative function (o.g.f.).}
\item{Build the algebraic grammar corresponding to the pseudoknots decomposition and deduce $F_2$ the o.g.f.}
\item{If $F_1 = F_2$ the decomposition is unambiguous.}
\end{itemize}
\end{comment}

\subsection{Simple-type pseudoknots}

\begin{figure}[t]
  {\centering \includegraphics[width=\textwidth]{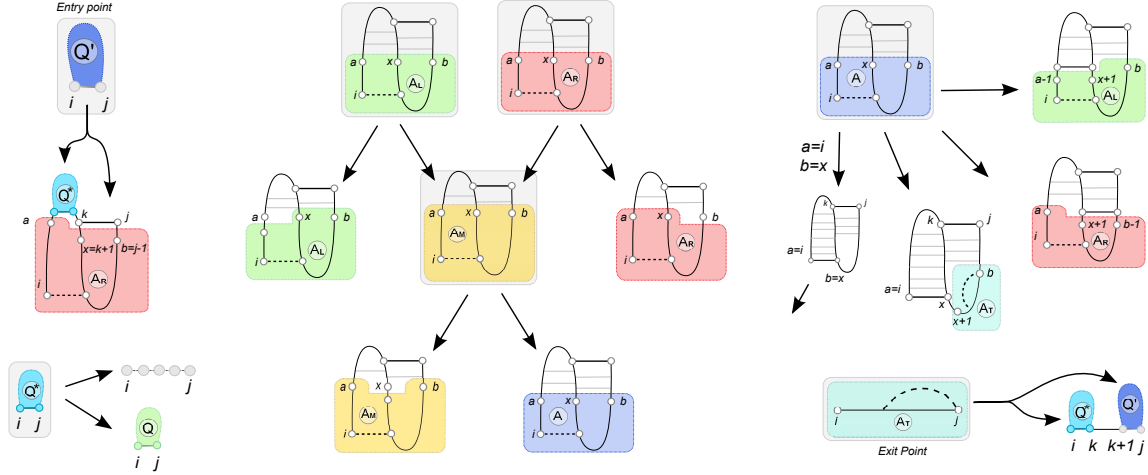}\\}
  \caption{An unambiguous decomposition for simple non-recursive pseudoknots that captures the Akutsu/Uemura class of pseudoknots. This decomposition yields $\BigO{n^4}/\Theta{(n^4)}$ time/memory algorithms for partially recursive pseudoknots and can be extended to include recursive pseudoknots and/or Turner energy contributions in $\BigO{n^5}/\Theta{(n^4)}$.\label{recursiveA}}
  \end{figure}

  In his seminal work, Akutsu~\cite{Akutsu2000} focused on a subset of pseudoknots motifs,
  the simple-type pseudoknots,
  and proposed algorithms of complexity in $\BigO{n^4}$ for simple non-recursive pseudoknots in a basepair-maximisation
  energy model, and in $\BigO{n^5}$ for recursive pseudoknots and loop-based energy models. However, the decomposition proposed in~\cite{Akutsu2000} is
  {\bf ambiguous}, e.g. there exists different ways to create unpaired regions. Therefore we propose in Figure~\ref{recursiveA} an unambiguous decomposition for the same conformation space.

  \ParHead{Previous results.} In a previous work~\cite{Saule2010,SaReStDe}, one of the authors showed that simple-type pseudoknots can be encoded by a simple formal language, in bijection with a context-free language. Here we focus on partly recursive simple pseudoknots presented in Figure~\ref{recursiveA}. They can be encoded by a well-parenthesized word $p$ over two systems of parentheses $\{(\ox,\bx),(\oy,\by)\}$, respectively indicating the leftmost and rightmost basepairs in Figure~\ref{recursiveA}, and an unpaired character $\oc$ such that
\begin{equation}p=(\oc^*\ox)^n\, p'\,(\oy\,\oc^*)^{m_1}\,(\bx\,\oc^*)^{n_1}\,(\oy\,\oc^*)^{m_2}\,(\bx\,\oc^*)^{n_2}\cdots (\oy\,\oc^*)^{m_k}\,(\bx\,\oc^*)^{n_k-1}\,\bx\,p''\,\by\,(\oc^*\by)^{m-1} \label{eq:simplePK}\end{equation}
where $k$ is some integral value, $\sum_{i=1}^k n_i=n \geq 1$, $\sum_{i=1}^k m_i=m \geq 1$, and $p',p''$ are any two recursively-generated conformations.

\ParHead{Completeness.} Let us show that the decomposition in Figure~\ref{recursiveA} is complete, i.e. that any partially recursive pseudoknot
can be generated by the decomposition.

Let us initially focus on base-pairs and ignore unpaired bases. The smallest word within the language of Equation~\ref{eq:simplePK} is $\ox p' \oy \bx p'' \by$ which can be generated by applying the initial case ($Q\to A_L \to A_M \to A \rightsquigarrow p' \ldots \oy \ldots \by$) followed directly by the terminal case ($A \to A_T \rightsquigarrow \ox\,p'\,\oy\,\bx\,p''\,\by$).
Moreover through a sequence $A\to A_R \to A_M \to A$, one adds an outermost edge around the right part $\oy \ldots \by$. So through $m$ iterations of the sequence the decomposition generates any structure $\oy^{m_1} \ldots \by^{m_1}$. Similarly through a sequence $A\to A_L \to A_M \to A$ one adds an outermost edge around the left part $\ox \ldots \bx$, and after $n_1$ iterations any structure $\ox^{n_1} \ldots \bx^{n_1}$ is generated.
Since these two sequences can be combined and alternated (starting with the initial case and finishing with the terminal case), then the decomposition generates any word
\begin{equation}
  p = \ox^{n} \, p' \, \oy^{m_1} \, \bx^{n_1}
         \, \oy^{m_2} \, \bx^{n_2}
         \, \cdots \,
         \, \oy^{m_k} \, \bx^{n_k} \, p'' \,
         \, \by^{m}\by.
\end{equation}
For the recursive call $p'$, it is easily verified that $Q^*$ generates any (PK) structure.
For $p''$ it is worth mentioning that, at a base-pairing level, $A\to A_T$ (right base paired) and $A\to \emptyset$ cover all possible situations.

Arbitrary numbers of unpaired bases $\oc$ can also be inserted right before the opening $\ox$ of a leftward base pair (resp. after closure $\bx$ of a leftward base pair, after the opening $\oy$ of a right base pair and before the closure $\by$ of a right base pair) by repeatedly applying the $A_L \to A_L$ (resp. $A_M \to A_M$, $A_L \to A_L$ and $A_M \to A_M$) rule after adding a left (resp. right) base pair. Consequently any structure described by a word in Equation~\ref{eq:simplePK} can be generated by the decomposition.

\ParHead{Unambiguity.} Let us now address the unambiguity of the decomposition, using our approach based on generating functions.
Equation~\ref{eq:simplePK} immediately gives a system of equations relating $AU(z)$, the generating function of simple partially recursive pseudoknots,
to $S(z)$ the gen. fun. of all structures:
\begin{align*}
AU(z) &= \sum_{k\ge 1} \left(\frac{z}{1-z}\right)^n\, S(z)\,\left(\frac{z}{1-z}\right)^{m_1}\,\left(\frac{z}{1-z}\right)^{n_1}\cdots \left(\frac{z}{1-z}\right)^{n_k-1}z\, S(z)\,z\left(\frac{z}{1-z}\right)^{m-1} = \frac{z^4\,S(z)^2\,(1-z)}{1-2\,z-z^2}.
\end{align*}
Now consider the dynamic programming decomposition illustrated by Figure~\ref{recursiveA}. Associating generating functions to each type of vertices and translating assigned bases into monomials, we obtain the following system of equations:
\begin{align*}
Q'(z) & = z^2\,S(z)\,A_R(z) &
A_L(z) & = z\,A_L(z) + A_M(z)&
A_R(z) & = z\,A_R(z) + A_M(z)\\
A_M(z) & = z\,A_M(z) + A(z)&
A(z) & = z^2\,A_R(z) + z^2\,A_L(z) + z^2\,S(z)& A_T(z) & = S(z)(1-z)-1.
\end{align*}
The last expression for $A_T(z)$ follows directly from the observation that any structure in $Q$ can be written
as a sequence of structures from $Q'$ interleaved with sequences of unpaired bases. Given that $A_T$ cannot feature
unpaired bases on its right end, one of the sequence of unpaired base must be removed. Furthermore $A_T$ does not generate
the empty structure, so we have $S(z) = (A_T(z)+1)/(1-z)$. Solving the system gives $ Q'(z) = \frac{S^2(z)\,z^4\,(1-z)}{1-2\,z+z^2} = AU(z)$
and the unambiguity/correctness of the decomposition directly follow.

\subsection{Fully-recursive kissing hairpins}
\begin{figure}[t]
\centering
\includegraphics[width=\textwidth]{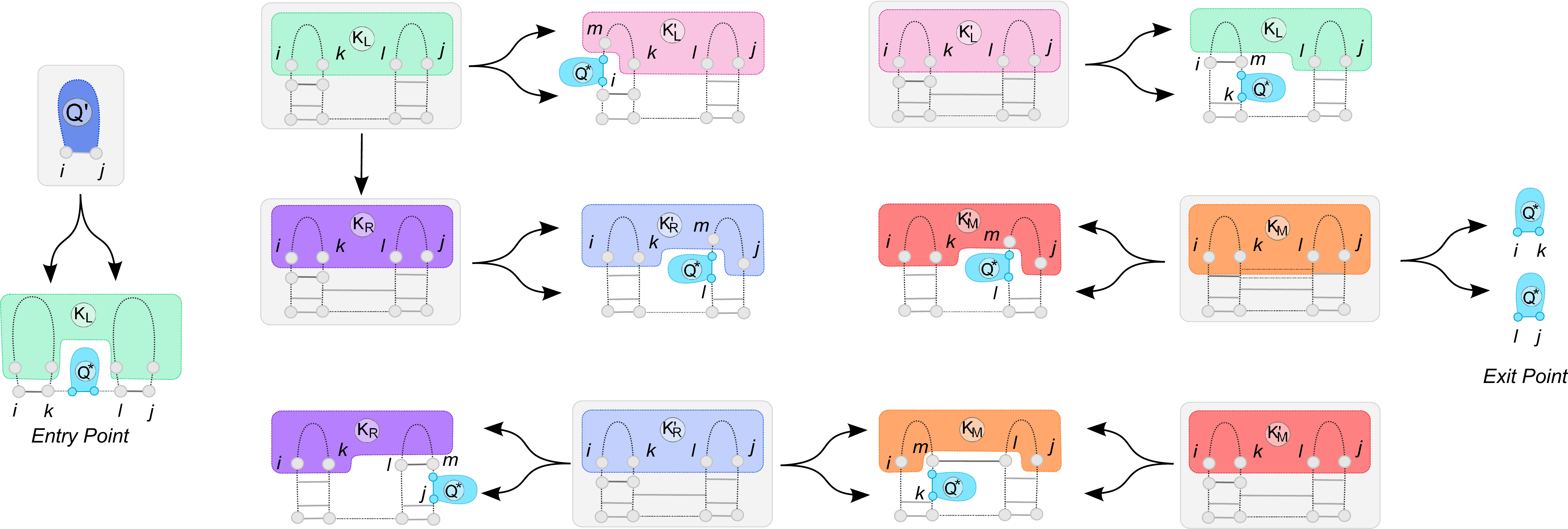}
\caption{Unambiguous decomposition of fully recursive kissing hairpins.}\label{recursiveKH}
\end{figure}
Kissing hairpins (KH) are pseudoknotted structure composed of two helices whose terminal loops are linked by a third helix.
These pseudoknots are frequently observed, and are exhaustively predicted by Chen~\emph{et al}~\cite{Chen2009} in time complexity in $\BigO{n^5}$, and in $\BigO{n^3}/\BigO{n^4}$ under restrictions by Theis~\emph{et al}~\cite{Theis2010}. Figure~\ref{recursiveKH}  presents an unambiguous decomposition which generates the space of recursive kissing hairpins.

\ParHead{Previous results.} Again, an encoding of kissing hairpins can be found in earlier work by one of the authors~\cite{Saule2010}, showing that any KH pseudoknot can be represented by a word $p$ over three systems of parentheses $\{(\ox,\bx),(\oy,\by),(\oz,\bz)\}$ (respectively denoting leftmost, central and rightmost helices) such that:
\begin{equation} p = (\ox S)^{n} \, (\oy S)^{m} \, (\bx S)^{n} \, (\oz S)^{k} \, (\by S)^{m} \, (\bz S)^{k-1} \, \bz.\label{eq:languageKH}\end{equation}
\ParHead{Completeness.} First let us remark that the \emph{minimal} conformation generated by the decomposition is $K_L \to K_R \to K'_{R} \to K_M \rightsquigarrow \ox S\oy S\bx S\oz S \by S \bz$. Remark that one can iterate arbitrarily over the states
$K_L \to K'_{L} \to K_L$, $K'_{R} \to K_{R} \to K'_{R}$ and $K'_M \to K_M \to K_M$.
Consequently one may \emph{insert} patterns $(K_L \to K'_{L} \to K_L)^{n-1} \rightsquigarrow (S\,\ox)^{n-1} \cdots (\bx\,S)^{n-1}$,
$(K'_{R} \to K_{R} \to K'_{R})^{k-1} \rightsquigarrow (\oz\,S)^{k-1} \cdots (\bz\,S)^{k-1}$ and $(K_M \to K'_M  \to K_M)^m \rightsquigarrow (\oy\,S)^{m-1} \cdots (S\,\by)^{m-1}$ in the minimal word above, and produce any conformation denoted by
$$\ox(S\ox)^{n-1} S(\oy S)^{m-1} y S(\bx S)^{n-1} \bx S \oz S (\oz S)^{k-1}\by(S \by)^{m-1} S (\bz S)^{k-1} \bz$$
where one recognizes the language of Equation~\ref{eq:languageKH} upon simple expansion.

\ParHead{Unambiguity.}
Equation~\ref{eq:languageKH} allows to derive the generating function $KH(z)$ of kissing-hairpin as a function of $S(z)$ the gen. fun. of all structures:
\begin{equation}
  KH(z) = \sum_{n,m,k \ge 1} (z S(z))^{n} (z S(z))^{m} (z S(z))^{n} (z S(z))^{k} (z S(z))^{m} (z S(z))^{k-1} z = \frac{z^6S(z)^5}{(1-z^2 S(z)^2)^3}\cdot
\end{equation}
Now consider the dynamic programming decomposition illustrated by Figure~\ref{recursiveKH}, and translate it into a system of functional equation:
\begin{align*}
   && K(z) & = z^4 K_L(z)S(z) & &\\
   K_L(z) & = S(z) K_L'(z) + K_R(z)&
   K_L'(z) & = z^2 K_L(z) S(z)&
   K_M(z) & = K_M'(z)S(z) + S(z)^2\\
   K_M'(z) & = z^2K_M(z)S(z) &
   K_R(z) & = K_R'(z) S(z)&
   K_R'(z) & = z^2 K_R(z) S(z) + z^2 K_M(z) S(z)
\end{align*}
Solving the system gives $K(z) = \frac{z^6S(z)^5}{(1-z^2 S(z)^2)^3} = KH(z)$ and the unambiguity of the decomposition immediately follows. Again hypergraphs algorithms can be used, and specialize into the complexities summarized in Table~\ref{tab:recapAlgo}.